\documentclass[twocolumn,showpacs,preprintnumbers,amsmath,amssymb,superscriptaddress,nofootinbib,pra]{revtex4-1}
\usepackage[dvipdfmx]{graphicx}
\usepackage{bmpsize}
\usepackage{indentfirst}
\usepackage{graphicx}
\usepackage{amsmath}
\usepackage{epsfig}
\usepackage{bm}
\usepackage{cases}
\usepackage{empheq}
\usepackage[usenames]{color}
\usepackage{here}

\def\>{{\rangle}}
\def\<{{\langle}}

\begin{document}
\title{ A new perspective on the Fano absorption spectrum in terms of  complex spectral analysis}

\author{Taku \surname{Fukuta}}
\affiliation{Department of Physical Science, Osaka Prefecture
University, Gakuen-cho 1-1, Sakai 599-8531, Japan}

\author{Savannah  \surname{Garmon}}
\affiliation{Department of Physical Science, Osaka Prefecture
University, Gakuen-cho 1-1, Sakai 599-8531, Japan}

\author{Kazuki \surname{Kanki}}
\affiliation{Department of Physical Science, Osaka Prefecture
University, Gakuen-cho 1-1, Sakai 599-8531, Japan}

\author{Ken-ichi \surname{Noba}}
\affiliation{Department of Mathematical Science, Osaka Prefecture
University, Gakuen-cho 1-1, Sakai 599-8531, Japan}

\author{Satoshi \surname{Tanaka}}
\email{stanaka@p.s.osakafu-u.ac.jp}
\affiliation{Department of Physical Science, Osaka Prefecture
University, Gakuen-cho 1-1, Sakai 599-8531, Japan}

\begin{abstract}
\today
\vspace{5pt}

A new aspect of understanding a Fano absorption spectrum  is presented in terms of the complex spectral analysis.
The absorption spectrum of an impurity embedded in semi-infinite superlattice is investigated.
The boundary condition on the continuum causes a large energy dependence of the self-energy, enhances the nonlinearity of the eigenvalue problem of the effective Hamiltonian, yielding several nonanalytic resonance states.
The overall spectral features is perfectly reproduced by the direct transitions to these discrete resonance states.
Even with a single optical transition path the spectrum exhibits an asymmetric Fano profile, which is enhanced for the transition to the nonanalytic resonance states.
Since this is the genuine eigenstates of the total Hamiltonian, there is no ambiguity in the interpretation of the absorption spectrum, avoiding the arbitrary interpretation based on the quantum interference.
The spectral change around the exceptional point is well understood when we extract the resonant state component.

\end{abstract}

\maketitle

\section{Introduction}

The Fano effect is a ubiquitous phenomena in quantum mechanics, recognized as a manifestation of  quantum interference \cite{Miroshnichenko10}.
In his seminal paper, Fano revealed that the absorption spectrum  in the photoionization process of an inner-shell electron shows a characteristic asymmetric spectral profile  known as {\it the Fano profile}\cite{Beutler35,Fano35,Fano61}, as a result of the interference between the direct photoionization transition and the ionization transition mediated by a resonance state.
Since then, a growing number of works have been devoted to study the Fano resonance in various physical systems.
However, it has been realized that there are cases that cannot be simply fit to the original interpretation of the interference of  multiple transitions to a common continuum.
As an example, the absorption spectrum in a quantum well shows a distinct Fano resonance, even though there is no direct transition to the continuum \cite{Faist97Nature}. 

In order to explain the Fano profile for a broader range of physical situations, the phenomenological effective Hamiltonian has been proposed \cite{Paspalakis99PRL}, which is represented by a finite non-Hermitian matrix of complex constants.
The discrete resonance eigenstates are identified as the eigenstate of the effective Hamiltonian with complex eigenvalues whose imaginary part represents the decay rate of the resonance state.
While the idea of the resonance state may go back to the early days of the study of nuclear reactions where the resonance state has been obtained by Feshbach projection method \cite{Feshbach62,Barz77NuclPhys}, there has recently been much focus on expanding the horizon of  quantum mechanics so as to incorporate the irreversible decay process directly into  quantum theory\cite{Hatano96PRL,Bender98PRL,Bender03,Heiss04JPA,Rotter09JPhysA,Rotter15,MoiseyevBook}.
In these studies, the starting Hamiltonian itself is taken as a non-Hermitian or  a Parity-Time-symmetry (PT-symmetry) Hamiltonian at the beginning\cite{Bender98PRL,Bender03}.

Even though the non-Hermitian effective Hamiltonian is useful to reproduce the Fano spectral profile, it is not clear how these matrix elements have been derived from time-reversible microscopic dynamics.
Indeed, since the derivation of the effective Hamiltonian usually counts on the Weisskopf-Wigner approximation \cite{Fountoulakis06PPRA}, i.e. Markovian approximation, the validity of the effective Hamiltonian should be reexamined.
In addition, the microscopic information of the interaction with the continuum is missing because the effect of the interaction is rewritten into complex constants that are phenomenologically determined.

On the other hand, there have been efforts to derive  a non-Hermitian effective Hamiltonian from the  microscopic total (Hermitian) Hamiltonian with use of the Brillouin-Wigner-Feshbach projection operator method (BWF method)\cite{Feshbach62,Jung99,Rotter09JPhysA,Rotter15,Rotter10,Eleuch15}, where  the effect of the  microscopic interaction with the continuum is represented by the {\it  energy dependent} self-energy function.
Prigogine and one of the authors (T.P.) {\it et al.} have clarified that the spectrum of the effective Hamiltonian coincides with that of the total Hamiltonian, revealing that the total Hermitian Hamiltonian may have complex eigenvalues due to the resonance singularity if we extend the eigenvector space from the ordinary Hilbert space to the extended Hilbert space, where the Hilbert norm of the eigenvector vanishes\cite{BohmBook,Petrosky91Physica,Petrosky97}.
It should be emphasized that the complex eigenvalue problem of the effective Hamiltonian becomes nonlinear in this approach in the sense that the effective Hamiltonian depends on its eigenvalue.
It has been revealed recently that this nonlinearity of the eigenvalue problem of the effective Hamiltonian causes interesting phenomena, such as the dynamical phase transition\cite{Tanaka16PRA}, bound states in continuum (BIC)\cite{Tanaka07,Garmon09}, as well as non-analytic spectral features \cite{Garmon12,Garmon15,Kanki17JMP}, and modified time evolution near exceptional points\cite{Garmon17JMP}.

Our aim in this paper is to show that the Fano absorption profile is well explained in terms of the complex spectral analysis, even  in the absence of multiple transition paths.
As a typical system, we consider the core-level absorption of an impurity atom embedded in a semi-infinite superlattice, where the charge transfer decay following the optical excitation is reflected in the absorption spectrum  (see Fig.\ref{Fig:Model}).
The Existence of the boundary with an infinite potential wall at the end of the chain causes striking effects on the resonance states in contrast to the infinite chain case: The self-energy has a strong energy dependence over the entire energy range of the continuum, which enhances the nonlinearity of  the eigenvalue problem of the effective Hamiltonian.
As a result, even with a single  impurity state, there appear several discrete resonance states that are nonanalytic in terms of the coupling constant.

We have discovered that the absorption spectrum is essentially determined by a sum of the direct transitions to these discrete resonance states.
The spectral profile due to the transition  takes an asymmetric Fano shape, even when only a single optical transition channel is present.
This is because the transition strength (oscillator strength), which is ordinarily real valued \cite{LoudonBook}, becomes complex as a result of the fact that the resonance state of the total Hamiltonian belongs to the extended Hilbert space.
Since our interpretation is based on the genuine eigenstate representation of the total Hamiltonian,  there is no ambiguity in the interpretation of the origin of the asymmetry in the absorption spectrum, contrary to the method using quantum interference in terms of the Hilbert space basis.

We have also found that because of the nonlinearity in the effective Hamiltonian there appears exceptional points (EP) where two resonance states coalesce  in terms of  not only energies but also 
their eigenstates\cite{KatoBook}.
We reveal that the absorption spectrum around the EP shows a broad single peak structure consisting of absorption transitions to the nearby degenerate nonanalytic resonance states.

The paper is  organized as follows.
We introduce our model for a semi-infinite chain including a two-level impurity atom incorporating a single intra-atomic optical transition in Section \ref{Sec:Model}.
The complex eigenvalue problem of the total Hamiltonian is solved in Section \ref{Sec:cEVP} and we present the characteristic behaviors of the trajectories of  the eigenvalues of the effective Hamiltonian in the complex energy plane. 
The absorption spectrum is studied in terms of complex spectral analysis in Section \ref{Sec:Absorption}, which is followed by some concluding discussions in Section \ref{Sec:Discussion}.

 \section{Model}\label{Sec:Model}
 
\begin{figure}[t]
  \begin{center}
   \includegraphics[height=5cm,width=8.5cm]{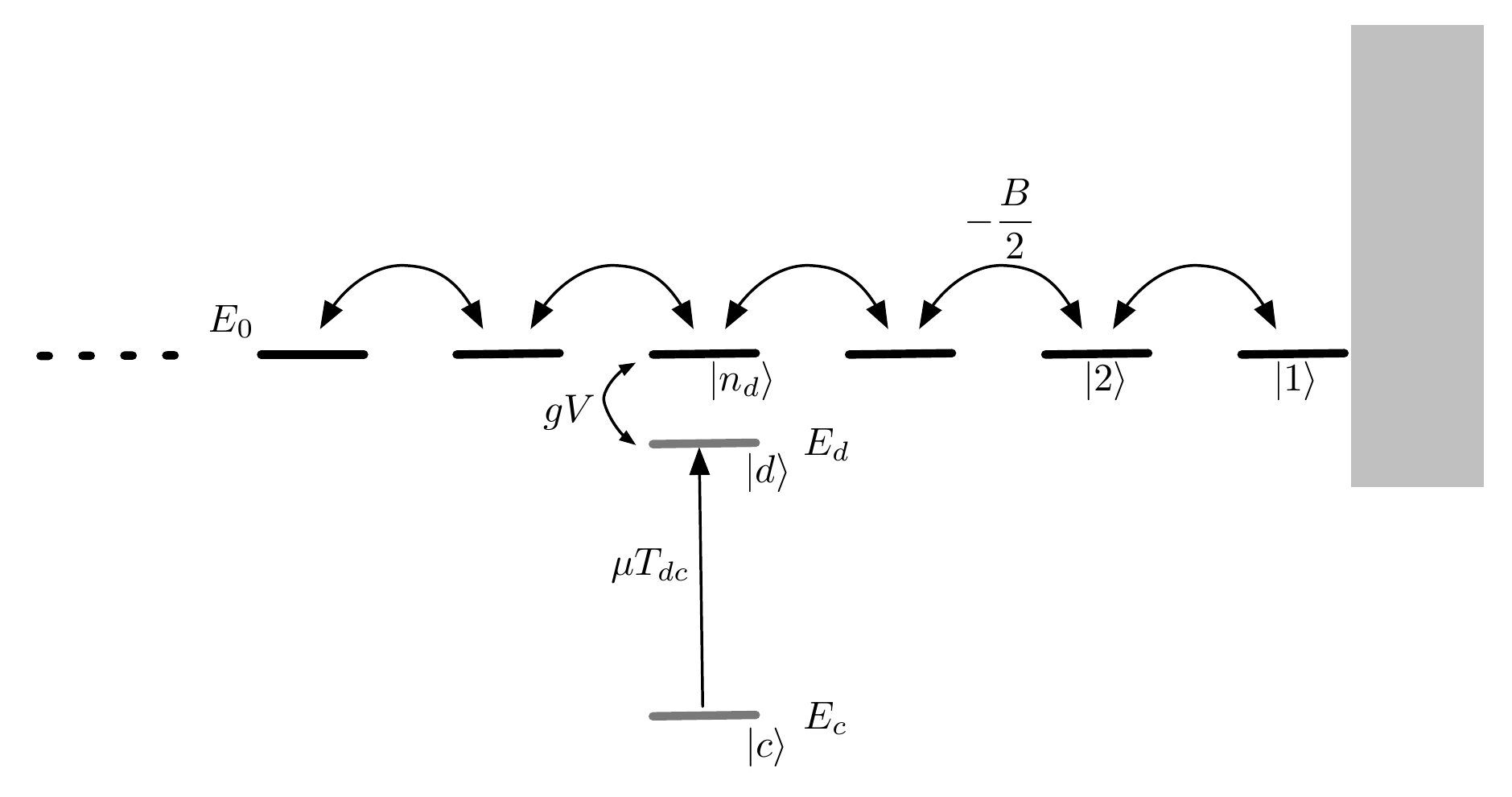}
\caption{The semi-infinite tight-binding chain with  a two level-impurity atom at  the $n_d$-th site from the boundary.
}
  \label{Fig:Model}
  \end{center}
\end{figure}
 
Our model consists of a  semi-infinite one-dimensional  superlattice with a two-level impurity atom  as shown in Fig.\ref{Fig:Model}, where the two-level atom is located at a distance $n_d \,a$ from the boundary.
In this work we take the lattice constant $a$ as our unit length, i.e. $a=1$.
We first consider a finite chain with the length $N$, then the Hamiltonian reads
\begin{align}\label{Hts}
\hat{H} &= E_{c} | c \>\<c | + E_{d} | d \>\<d | +\sum_{n=1}^N E_0|n\>\<n| \notag\\
& - \frac{B}{2} \sum_{n=1}^{N-1} \left( | n+1\>\<n | + {\rm H.c. }\right)+ gV  \left(  | n_d \>\<d | + {\rm H.c.}  \right).
\end{align}
The wavenumber state for the semi-infinite lattice is defined by
\begin{align}\label{ksemi}
| k_j \> \equiv \sqrt{2\over  N}\sum_{n=1}^N   \sin( k_j n) |n \>\;,
\end{align}
where $k_j$ takes $k_j=\pi j/ (N+1) \; (j=1, \cdots, N)$ under the fixed boundary condition.
In the limit $N\to \infty$, the discrete variable  $k_j$ becomes  continuous  in  $0< k <\pi$, and the summation becomes an integral over $k$.
Applying  the transform for the discrete wave number  $k_j$ to the continuous $k$ in the limit $N\to \infty$,  defined by\cite{Petrosky97}
\begin{equation}
|k\> \equiv \left({N\over  \pi}\right)^{1/2} |k_j\>  \;,
\end{equation}
the continuous unperturbed basis $|k\>$ satisfies the orthonormality according to  Dirac's delta function
\begin{equation}
\<k|k'\>=\delta(k-k')
\end{equation}
and the completeness relation 
\begin{equation}
1=|d\>\<d|+\int_0^\pi |k\>\<k| dk \;.
\end{equation}
With use of this basis,  the total Hamiltonian is rewritten as  
\begin{align}\label{Htk}
\hat{H}  = \hat H_0+g \hat V \;,
\end{align}
where
\begin{subequations}\label{AppEq:H0V}
\begin{align}
\hat H_0&=E_c |c\>\<c| + E_{d} | d \>\< d | + \int_0^\pi  E_{k} | k \>\< k |  dk   \;,\\
g\hat V&=\int_0^\pi gV_{k} \left(   | k \rangle \langle d | + |d\>\<k|  \right)  dk\;.
\end{align}
\end{subequations}
The energy dispersion of the continuum is given by
\begin{align}\label{Ek}
E_{k} =E_0-B \;{\rm cos} \;k \;,
\end{align}
and  the interaction potential  $V_k$ in terms of the continuous wave number  is given by
\begin{align}\label{SemVk}
V_k\equiv \left({2\over \pi}\right)^{1/2}V \sin(n_d k) \;.
\end{align}
Hereafter, we take $E_0=0$ and $B=1$, as the energy origin and the energy unit, respectively.

In this paper, we consider a single intra-atomic optical transition induced by  incident light with frequency $\omega$ under the dipole approximation.
The transition operator is given by
\begin{align}\label{T}
\hat{T} \equiv \mu \bigl( T_{dc} | d \rangle \langle c | + {\rm H.c.} \bigr),
\end{align}
with  a dimensionless coupling constant $\mu$, where we have adopted the rotating-wave-approximation (RWA) in the weak coupling case $\mu\ll 1$.

Using the first order time-dependent perturbation method for the interaction between  light and matter,  the absorption spectrum  is given  by \cite{Cohen-Tannoudji_2008,Tanaka06PRB}
\begin{align}\label{Fo}
F(\omega)&={1\over \pi}  {\rm Re}\int_0^\infty dt \, e^{i(\omega+E_c) t-\epsilon t}  \<c|\hat T e^{-i\hat H t}\hat T|c\> \notag\\
&=-{\mu^2 T_{dc}^2\over \pi}\;{\rm Im}\<d| {1\over \Omega-\hat H +i\epsilon } |d\>  \;,
\end{align}
where we denote $\Omega\equiv \omega+E_c$, and we have used Eq.(\ref{T}).\footnote{In Eq.(\ref{Fo}), we have dropped the factor $2\pi/\hbar$ for convenience.}

Even though the Green's function method yields  an analytical formula for $F(\Omega)$ for this system as shown in Appendix \ref{App:Green}, we shall present an alternative way to interpret the absorption profile in terms of  resonance states, which are considered to be decaying elementary excitations inherent to a given system.

\section{Complex eigenvalue problem } \label{Sec:cEVP}

We begin our analysis by solving the complex eigenvalue problem of $\hat H$ \cite{Petrosky91Physica}:
\begin{align}\label{EVH}
\hat{H} | \phi_\xi \> = z_\xi | \phi_\xi\>\;,\; \<\tilde{\phi}_\xi | \hat{H} = z_\xi \< \tilde{\phi}_\xi | \;,
\end{align}
where the right- and the left-eigenstates, $|\phi_\xi\>$ and $\<\tilde\phi_\xi|$, respectively, share the same eigenvalue $z_\xi$; we use a greek index for the (anti-)resonant states with complex eigenvalues and a roman index for the eigenstates (bound or continuum) with  real eigenvalues. 
These eigenstates satisfy  biorthonormality and  bicompleteness:
\begin{align}
\delta_{\xi,\xi'}=\<\tilde\phi_\xi|\phi_{\xi'}\> \;,\; 1=\sum_\xi |\phi_\xi\>\<\tilde\phi_\xi| \;.
\end{align}
In the present model, just as in the case that we studied in Ref.\cite{Tanaka06PRB}, the bicomplete basis set of the total Hamiltonian is composed of the discrete resonant states, the continuous state, as well as the stable bound states:
\begin{align}\label{biComp3}
 \sum_{i\in\rm{R}^{I}} | \phi_i \>\< \phi_i | +\sum_{\alpha=1}^{n_0-1} |\phi_\alpha\>\<\tilde\phi_\alpha|+ \int_0^\pi dk | \phi_k \>\< \tilde\phi_k  | = 1 \;,
\end{align}
where the first, the second, and the third  terms represent the bound states in the first Riemann sheet, the resonance states, and the continuous states, respectively.
Since this decomposition of the identity is represented by the eigenstates of the total Hamiltonian, it is  essential to understand the absorption spectra in terms of the irreversible decay process emerging from the time-reversible microscopic dynamics.

In order to obtain the discrete resonance states, 
we remove the infinite number of degrees from the problem  by using the Brilluoin-Wigner-Feshbach projection method via the projection operators \cite{Feshbach62}\footnote{Since the core level is decoupled from $|d\>$ and $|k\>$ states in $\hat H$, we here focus on the other terms than  the first one  in $\hat H_0$ given by Eq.(\ref{AppEq:H0V}a)}
\begin{align}\label{Projections}
\hat{P}^{(d)} \equiv | d \>\<d | \;, \; \hat{Q}^{(d)} \equiv 1 - \hat{P}^{(d)} =\int_0^\pi | k \>\< k | dk \;,
\end{align}
where $\hat P^{(d)}$ is the projection for the impurity state and $\hat Q^{(d)}$ is its complement.

We apply the projection operators  to the right-eigenvalue problem (the first equation of Eq.(\ref{EVH}) ), which then reads
\begin{subequations}\label{Projection_d}
\begin{align}
\hat P^{(d)} \hat H_0\hat P^{(d)} |\phi_\alpha\>+\hat P^{(d)} g\hat V\hat Q^{(d)} |\phi_\alpha\>&=z_d\hat P^{(d)}|\phi_\alpha\>  \;,\\
\hat Q^{(d)} g\hat V\hat P^{(d)} |\phi_\alpha\>+\hat Q^{(d)} \hat H\hat Q^{(d)} |\phi_\alpha\>&=z_d\hat Q^{(d)}|\phi_\alpha\> \;.
\end{align}
\end{subequations}
The $\hat Q^{(d)}|\phi_\alpha\>$ component is solved in Eq.(\ref{Projection_d}b) as
\begin{align}\label{AppEq:Qphi}
\hat Q^{(d)}|\phi_\alpha\>={1\over z_\alpha-\hat Q^{(d)} \hat H\hat Q^{(d)}}\hat Q^{(d)}\hat H\hat P^{(d)}|\phi_\alpha\> \;,
\end{align}
which, after substitution into Eq.(\ref{Projection_d}a),  gives the right-eigenvalue problem of the effective Hamiltonian 
\begin{align}\label{HeffEVP}
\hat H_{\rm eff}(z_\xi) \hat P^{(d)}| \phi_{\xi} \> = z_{\xi} \hat P^{(d)} | \phi_{\xi} \> \;, 
\end{align}
where the effective Hamiltonian $\hat H_{\rm eff}(z) $ is defined by
\begin{subequations}\label{Heff}
\begin{align}
\hat H_{\rm eff}(z)&=\hat P^{(d)}\hat H_0  P^{(d)} \notag\\
&\quad+  P^{(d)}\hat V\hat Q^{(d)} {g^2\over z-\hat Q^{(d)} \hat H \hat Q^{(d)}} \hat Q^{(d)} \hat V  P^{(d)}  \\
&=\left[E_d+g^2 \Sigma^+(z)\right]\hat P^{(d)}\;, 
\end{align}
\end{subequations}
and  the self-energy $\Sigma^+(z)$ is given by  
\begin{align}\label{Self}
\Sigma^+(z)&= \int_0^\pi dk {V_k^2 \over (z-E_k)^+} ={2\over \pi} \int_0^\pi dk {V^2 \sin^2 (n_d k) \over (z-E_k)^+} \notag\\
 &={V^2\over \sqrt {z^2-1}} \left[ 1-\left(z-\sqrt{z^2-1}\right)^{2n_d} \right] \;.
\end{align}
Note that   $\Sigma^+(z)$ is  defined by the Cauchy integral with the branch cut from -1 to 1 in the energy plane; we define $\Sigma^+(z)$ by taking the analytic continuation  from the upper half energy plane  as denoted by the $+$ superscript\cite{Petrosky91Physica}.
This is the same self-energy that brings about  the bound-state-in-continuum (BIC) as studied in Ref.\cite{Tanaka07}. (See Eq.(4) in \cite{Tanaka07}.)
In general, the self-energy has a strong energy dependence near the lower energy bound of the continuum, which results in  non-exponential quantum decay  \cite{Khalfin58JETP,Rothe06PRL}.
In addition, when the potential $V_k$ changes with the wave number $k$, the self-energy depends on the energy.
In the present system, the latter yields  oscillations in $\Sigma^+(z)$ with energy as shown in Fig.2 in Ref.\cite{Tanaka07}, in contrast to the infinite chain case \cite{Tanaka06PRB}.
In fact, with the change of  variables  $z=-\cos \theta$, the self-energy is written by
\begin{align}\label{Self2}
\Sigma^+(z)={V^2\over i \sin\theta}\left(1-e^{i 2n_d \theta} \right) \;,
\end{align}
leading to the BIC points 
\begin{align}\label{BIC}
E_k^{\rm BIC}=-\cos\left({\pi k \over n_d}\right)  \quad (k=1,\cdots,n_d-1) \;,
\end{align}
where the wave function is  confined in between the boundary and the impurity atom \cite{Tanaka07}.

Note that Eq.(\ref{HeffEVP})  is {\it nonlinear} in the sense that the operator itself depends on its eigenvalue due to the energy dependance of the self-energy as pointed out in the Introduction\cite{Petrosky91Physica,Tanaka16PRA,Kanki17JMP}. 
It is only when taking into account this nonlinearlity that the spectrum of the effective Hamiltonian $\hat H_{\rm eff}$ coincides with that of the total Hamiltonian $\hat H$, and the effective problem is dynamically justified.
Thus, the dispersion equation for $\hat H_{\rm eff}(z)$ reads
\begin{align}\label{DispersionSemi}
&\eta^+(z)\equiv z-E_d -g^2\Sigma^+(z)\notag\\
&=z-E_d- g^2{V^2 \over\sqrt {z^2-1}} \left\{ 1-\left( z-\sqrt{z^2-1}\right)^{2 n_d} \right\}=0 \;,
\end{align}
yielding $n_d+1$ discrete solutions, among which there are $n_d-1$ resonance state solutions with a negative imaginary part of the eigenvalues and the two real-valued  eigenvalues.
Here the boundary condition of  the infinite potential wall at the one end introduces strong  wave number dependence, which allows several discrete resonance states to appear such that the numbers of the resonance states increases as $n_d$ increases.

\begin{figure}
\begin{center}
\includegraphics[height=14cm,width=8.5cm]{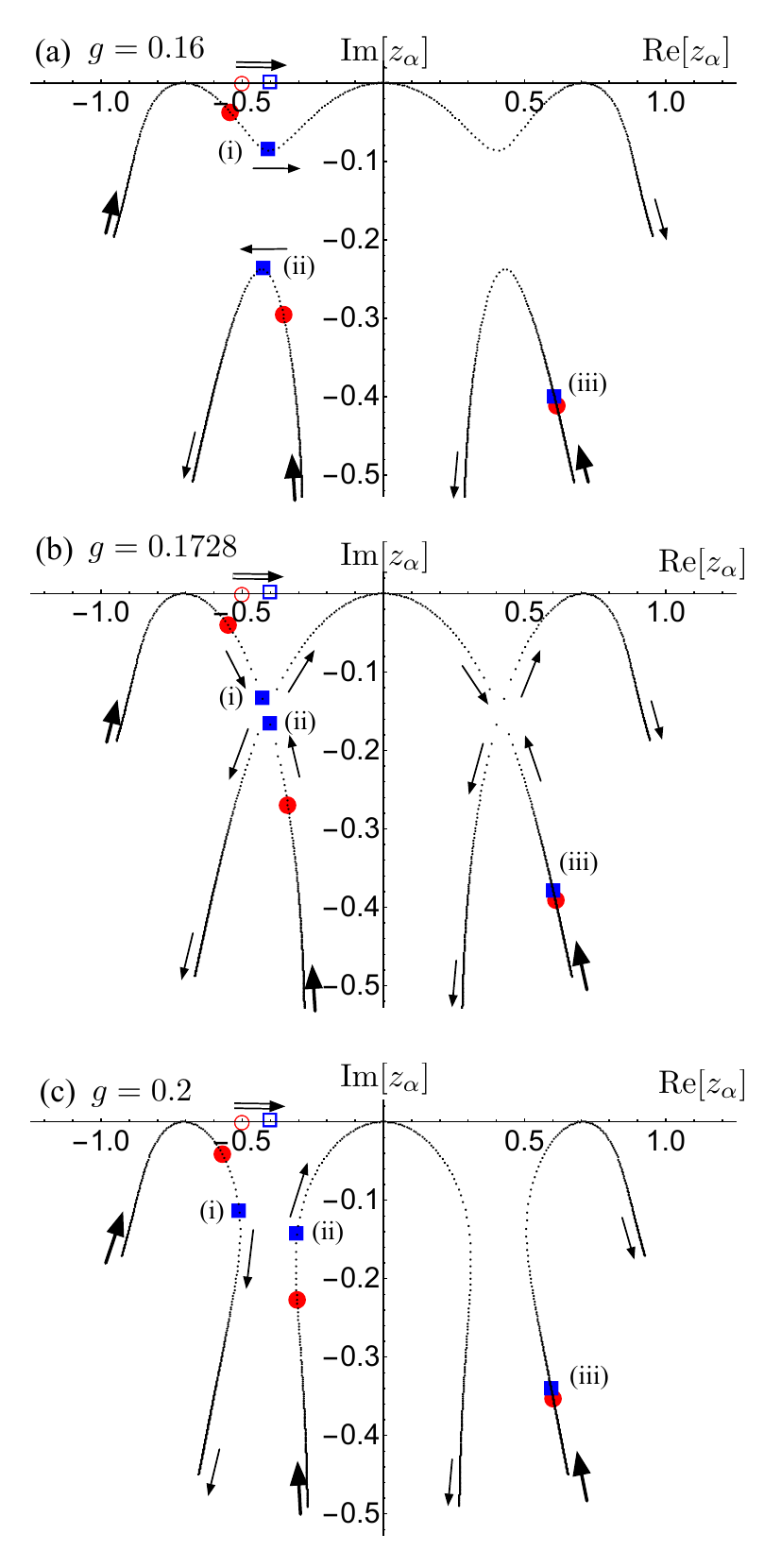}
\caption{(Color online) Resonance state solutions $z_\alpha$ of  Eq.(\ref{DispersionSemi}) for (a) $g=0.16$, (b) $g=0.1728\simeq g_{\rm EP}$, and (c)  $g=0.2$ for $n_d=4$ in the complex energy plane, where the horizontal and the vertical axes denote the real and the imaginary parts of the eigenvalues.
The filled (red) circles and (blue) squares denote the discrete resonance states for $E_d=-0.5$ and $E_d=-0.4\simeq E_{\rm EP}^-$, respectively, while the open ones denote the corresponding bare impurity energies $E_d$.
The dotted lines are the trajectory of the resonance state solutions with the change of $E_d$.
The arrows indicate the direction of the trajectories as $E_d$ increases as shown by the double arrows, and thick arrows indicate the entry at $E_d=-1$.
}
\label{Fig:ResonanceStates}
\end{center}
\end{figure}
As an illustration, we show the discrete resonance state solutions of Eq.(\ref{DispersionSemi}) for the  case  $n_d=4$ in the complex energy plane in Fig.\ref{Fig:ResonanceStates}, where the results are shown for (a)  $g=0.16$, (b) $g=0.1728$, and (c) $g=0.2$. 
There are three BIC points at $E_d=\pm 1/\sqrt{2}$ and  $0$, irrespective of the values of $g$, as seen from Eq.(\ref{BIC}).

We show the three resonance states for $E_d=-0.5$ and those for $E_d=-0.4$ by the filled circles and squares, respectively, while the corresponding bare impurity energies are also plotted by the open circles and squares on the real axis, respectively.
Hereafter we use the index $\alpha$ to distinguish the discrete resonance states.
The position of the three resonance states  change with $E_d$ as depicted by the dotted lines, which we shall call the {\it trajectories} of the resonance states.
In Fig.\ref{Fig:ResonanceStates}, we show the three trajectories as $E_d$  changes from $-1$ to 1.
The arrows indicate the directions of the trajectories with increasing of $E_d$, where the entries of the three trajectories from $E_d< -1$ are indicated by the thick arrows.

It is found that the feature of the trajectories are characteristically different for the three cases, distinguished at the critical value  $g_{\rm EP}=0.1728\cdots$, which corresponds to the exceptional point.
The exceptional point is a special point in  parameter space where not only the complex eigenvalues but also the wave functions coalesce  \cite{KatoBook}.
The exceptional points are obtained by looking for  parameter values satisfying the double root condition for the dispersion equation\cite{Garmon12,MoiseyevBook}, which requires in addition to Eq.(\ref{DispersionSemi}) that
\begin{align}\label{resultant}
{d\over dz}\eta^+(z;E_d,g)=1-g^2{d\over dz}\Sigma^+(z)=0 \;.
\end{align}
As shown in Fig.\ref{Fig:ResonanceStates}(b), we numerically obtained an exceptional point at $E_{\rm EP}^\pm=\pm0.3981\cdots$ for  $g_{\rm EP}\simeq 0.1728\cdots$, which is in between the BIC points for $E_d$:  $E_1^{\rm BIC}=-1/\sqrt{2}< E_{\rm EP}^-< E_2^{\rm BIC}=0$.
\footnote{There are several other exceptional points satisfying the double root conditions of  Eqs.(\ref{DispersionSemi}) and (\ref{resultant}). 
But in this work we focus on the spectral change when we change $E_d$ from one of the BIC points to the other: from $E_1^{\rm BIC}=-1/\sqrt{2}$ to $E_2^{\rm BIC}=0$.}

For $g=0.16(<g_{\rm EP})$ in Fig.\ref{Fig:ResonanceStates}(a), the trajectory numbered by (i)  is continuously close to the real axis (except near the band edge) which may be  identified as the perturbed solution of Eq.(\ref{DispersionSemi}).
On the other hand, the other two trajectories numbered by (ii) and (iii) are separated from the trajectory  (i).
These two solutions are  nonanalytic in terms of $g$: Indeed, the imaginary part of their complex eigenvalues are the order  $\gtrsim O(g) > O(g^2)$ for $g <  1$. 
As $E_d$ increases from $E_d=E_1^{\rm BIC}=-1/\sqrt{2}$ the trajectories of (i) and (ii)  come close, and around $E_d\simeq E_{\rm EP}^-$ they repel each other in parallel  to  the imaginary axis while moving in opposite directions as indicated by the arrows.
A similar repulsion happens between the trajectories (i) and (iii) around $E_d\simeq E_{\rm EP}^+$.

In the case of  $g=0.2(>g_{\rm EP})$ shown in Fig.\ref{Fig:ResonanceStates}(c), the three trajectories instead repel each other in parallel  to the real axis while traveling in opposite directions, and there is no continuous trajectory close to the real axis.
In this case, as $E_d$ increases from $E_d=E_1^{\rm BIC}=-1/\sqrt{2}$,  the trajectory (i) starts from the BIC value and shifts away from the real axis, and becomes strongly nonanalytic, while the trajectory (ii) moves in the  opposite direction.
The case  $g\simeq g_{\rm EP}$ shown in Fig.\ref{Fig:ResonanceStates}(b) is the boundary between these two cases.
The trajectories (i) and (ii) coalesce at $E_d=E_{\rm EP}^-=-0.3981\cdots$.
As will be shown in the next section, it is found that the characteristic difference of these behaviors in the trajectories are clearly reflected in the absorption spectrum.

Once we have solved the eigenvalue problem of the effective Hamiltonian, it is straightforward to obtain the complex eigenstate of the total Hamiltonian,
by adding the complement component.
The explicit representation of the right-eigenstate is obtained as
\begin{equation}\label{rightES}
|\phi_\alpha\>=\<d|\phi_\alpha\>\left( |d\>+g   \int_0^\pi dk {V_k \over (z-E_k)^+_{z=z_\alpha}} |k\>\right) \;,
\end{equation}
where the $+$ sign in the integrand indicates taking the analytical continuation from the upper energy plane to the resonance pole $z_\alpha$ as mentioned above.
The left-eigenstate is similarly obtained as
\begin{equation}\label{leftES}
\<\tilde\phi_\alpha|=\<\tilde\phi_\alpha| d\> \left( \<d|+g   \int_0^\pi dk {V_k \over (z-E_k)^+_{z=z_\alpha}} \<k|\right) \;.
\end{equation}
Note  the analytic continuation should be taken in the same direction as $|\phi_\alpha\>$, resulting in
\begin{equation}
\left(|\phi_\alpha\>\right)^\dagger \neq \<\tilde\phi_\alpha|  \;.
\end{equation}

The normalization condition for these eigenstates is given by
\begin{equation}
\<\tilde\phi_\alpha|\phi_{\alpha'}\>=\delta_{\alpha,\alpha'} \;,
\end{equation}
so that
\begin{equation}\label{Norm}
\<\tilde\phi_\alpha|d\>\<d|\phi_\alpha\>=\left( 1-g^2 {d\over dz}\Sigma^+(z)\Big|_{z=z_\alpha} \right)^{-1} \;.
\end{equation}
Taking a derivative of $z_\alpha(E_d)$ as a function of $E_d$ in Eq.(\ref{DispersionSemi}), we find \cite{Tanaka16PRA}
\begin{align}\label{NormRelation}
{d\over d E_d}z_\alpha(E_d)=\<\tilde\phi_\alpha|d\>\<d|\phi_\alpha\> \;.
\end{align}
It is clear that the normalization constant can be complex, which is rigorously determined by the normalization condition for the eigenstate of the {\it total} Hamiltonian.
Note that the normalization constant diverges at the EP\cite{Garmon12,Kanki17JMP,Garmon17JMP}. 
As will be seen in the next section, this complex normalization constant determines the absorption spectral shape, while the complex eigenvalues determine the peak position and the spectral width.

By using the projection operator  $\hat P^{(k)}\equiv |k\>\<k|$ and its complement, we have similarly obtained  the right-eigenstate for the continuous state \cite{Petrosky91Physica} in terms of the continuous wave number as
\begin{align}\label{phikR}
|\phi_k\>=|k\>+{g V_k \over \eta_d^+(E_k)} \left( |d\>+g \int_0^\pi dk' {V_{k'}\over E_k-E_{k'}+i\epsilon }|k'\> \right) \;,
\end{align}
where we have defined the delayed analytic continuation of the Green's function as \cite{Petrosky91Physica}
\begin{align}
{1\over\eta^+_d(E_k)}\equiv {1\over \eta^+(E_k)}\prod_\alpha{ E_k-z_\alpha \over (E_k-z)^+_{z=z_\alpha}}  \;,
\end{align}
with the inverse of the Green's function $\eta^+(z)$ given by Eq.(\ref{DispersionSemi}).
The left-eigenstate for the continuous state is also given by
\begin{align}\label{phikL}
\<\tilde\phi_k|=\<k|+{g V_k \over \eta^-(E_k)} \left( \<d|+g \int_0^\pi dk' {V_{k'}\over E_k-E_{k'}-i\epsilon }\<k'| \right) \;.
\end{align}
In Eqs.(\ref{phikR}) and (\ref{phikL}), $\epsilon$ is a positive infinitesimal.
Together with the two bound eigenstates with real eigenvalues, we come to the decomposition of the identity as shown in Eq.(\ref{biComp3}).

 \section{Absorption spectrum}\label{Sec:Absorption}

Applying  the bi-completeness Eq.(\ref{biComp3}) to the absorption spectrum Eq.(\ref{Fo}), we obtain the representation of the absorption spectrum in terms of the complex eigenstates given in the preceding section as
\begin{align}\label{FoComSemi}
 &F(\Omega)=\mu^2 T_{dc}^2  \sum_{i\in\rm{R}^{I}} \left| \<d|\phi_i\>\right|^2 \delta( \Omega - E_i) \notag\\
& -{ \mu^2 T_{dc}^2 \over\pi}  {\rm Im} \left[  \sum_\alpha  {\<d|\phi_\alpha\>\<\tilde\phi_\alpha|d\> \over \Omega - z_\alpha+ i \epsilon}  +\int {\<d|\phi_k\>\<\tilde\phi_k|d\> \over \Omega - E_k+ i \epsilon} dk\right] \;,
\end{align}
where the first term is attributed to the bound states, while the second and the third terms are attributed to the resonance states and the continuous eigenstates, respectively,  which  are directly related to the Fano effect.

\begin{widetext}

\begin{figure}[t]
\includegraphics[height=15cm,width=14cm]{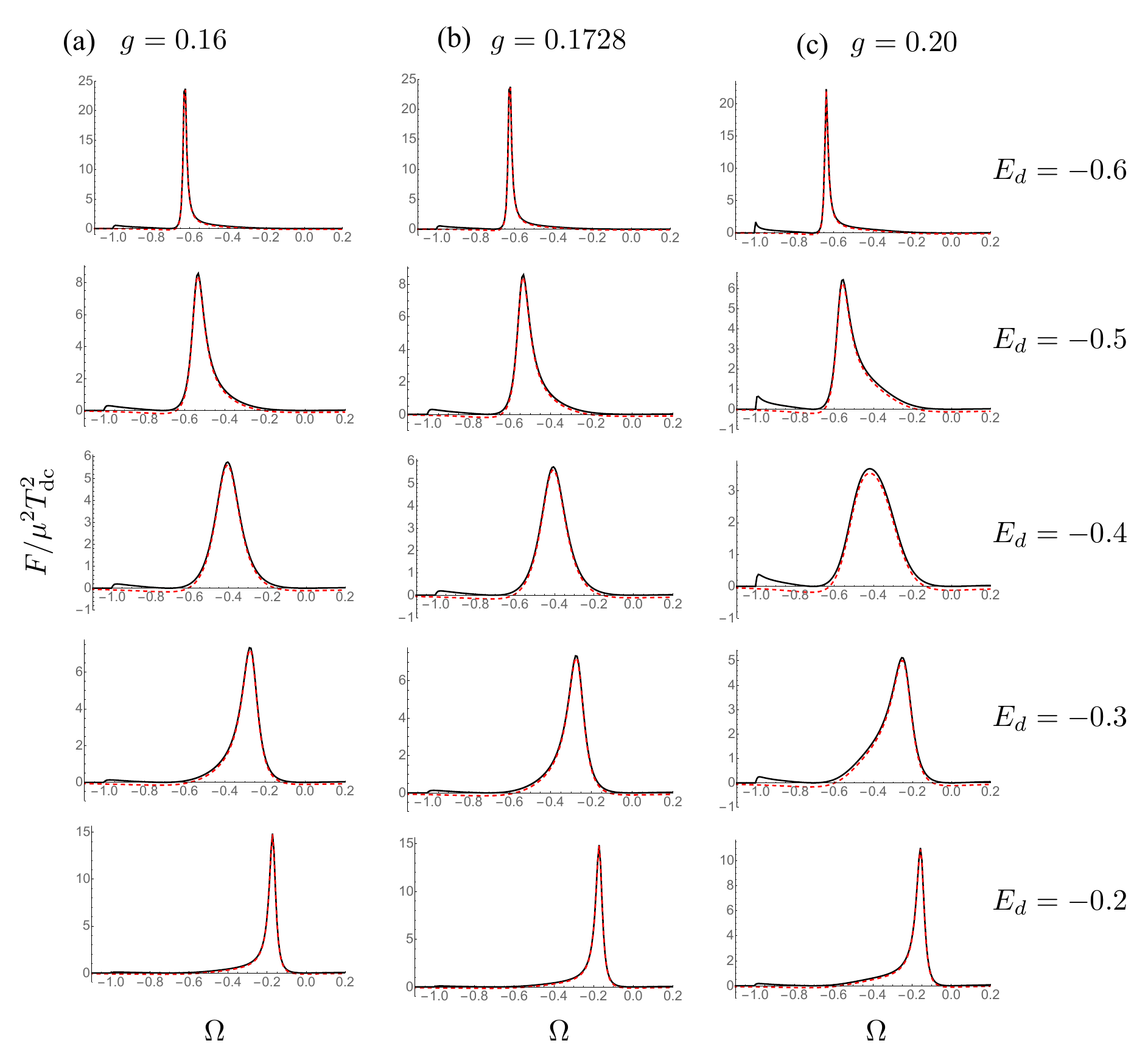}
\caption{(Color online) Absorption spectra $F(\Omega)$  corresponding to the case of Fig.\ref{Fig:ResonanceStates}: (a) $g=0.16$, (b) $g=0.1728$, and (c) $g=0.20$ for $n_d=4$, where the spectrum intensity is divided by $\mu^2 T_{\rm dc}^2$: $F(\Omega)/\mu^2 T_{\rm dc}^2$.
In each panel, we show the absorption spectra for $E_d=-0.6, -0.5, -0.4, -0.3$, and $-0.2$ from the top to bottom. 
The  (black)  solid lines represent $F(\Omega)$ and the (red) dotted lines represent the resonance state component of Eq.(\ref{FoComSemi}).
}
\label{Fig:absAll}
\end{figure}

\end{widetext}

We have shown in  Fig.\ref{Fig:absAll} the absorption spectra $F(\Omega)$ by the (black) solid lines and  the resonance state component by the (red) dotted lines,  corresponding to the case of Fig.\ref{Fig:ResonanceStates}: (a) $g=0.16$, (b) $g=0.1728$, and (c) $g=0.20$ for $n_d=4$.
In each panel, we show the absorption spectra for $E_d=-0.6, -0.5, -0.4, -0.3$, and $-0.2$ from  top to bottom.

It is striking that the absorption spectra are almost perfectly reproduced  only by the direct transitions to the discrete resonance states  given by the second term of Eq.(\ref{FoComSemi}).
We also find that the overall features of the spectral change are similar for the three cases: At $E_d=-0.6$ close to the BIC point $E_1^{\rm BIC}$, there is a sharp  peak with a high energy tail.
As $E_d$ increases, the peak position shifts  to the high energy side, and the spectrum is broadened. 
When $E_d$ comes close to $E_{\rm EP}$, the spectrum becomes a nearly-symmetric broad peak.
And as $E_d$ further increases toward the second BIC point $E_2^{\rm BIC}$, the peak position further shifts to the high energy side and it becomes a sharp peak  with a low energy tail.
Thus, even with a single intra-atomic optical transition, the absorption profile changes as $E_d$ changes.
It will be shown below that even though the overall features of the spectral changes are similar for the three cases, the spectral component is very different reflecting the characteristic difference in the complex eigenvalues in Fig.\ref{Fig:ResonanceStates}.

In order to demonstrate this, we decompose the resonant state component,  i.e., the second term of  Eq.(\ref{FoComSemi}), into  symmetric and antisymmetric parts according to the real and imaginary parts of the normalization constant $\<d|\phi_\alpha\>\<\tilde\phi_\alpha|d\>$ given by 
Eq.(\ref{NormRelation}).
The resonance state component for $|\phi_\alpha\>$  $(\<\tilde\phi_\alpha|)$  is written by
\begin{align}\label{fReso}
f_\alpha(\Omega)\equiv-{ \mu^2 T_{dc}^2 \over\pi}  {\rm Im} {\<d|\phi_\alpha\>\<\tilde\phi_\alpha|d\> \over \Omega - z_\alpha+ i \epsilon}= f^S_\alpha(\Omega)+ f^A_\alpha(\Omega) \;,
\end{align}
where 
\begin{subequations}\label{fSA}
\begin{align}
f_\alpha^S(\Omega)&\equiv{ \mu^2 T_{dc}^2 \over\pi}\cdot {\gamma_\alpha \over (\Omega-\varepsilon_\alpha)^2 +\gamma_\alpha^2}\cdot {d\varepsilon_\alpha\over d E_d} \;, \\
f_\alpha^A(\Omega)&\equiv{ \mu^2 T_{dc}^2 \over\pi}\cdot {(\Omega-\varepsilon_\alpha)  \over (\Omega-\varepsilon_\alpha)^2 +\gamma_\alpha^2} \cdot{d\gamma_\alpha \over d E_d} \;,
\end{align}
\end{subequations}
 and   $z_\alpha=\varepsilon_\alpha-i\gamma_\alpha$ $(\gamma_\alpha >0)$.

The first factors of Eqs.(\ref{fSA})  are the optical transition strengths.
The second factors determine the spectral profiles: symmetric and antisymmetric profiles for Eq.(\ref{fSA}a) and Eq.(\ref{fSA}b),  respectively, whose  maximum values are ${1/ 2\gamma_\alpha}$.
The factor $d\gamma_\alpha/dE_d$ in Eq.(\ref{fSA}b) determines  the degree of the Fano-type asymmetry.
In the weak coupling case under the Markovian approximation, where the energy dependence of the self-energy may be neglected, $\gamma_\alpha$ does not depend on $E_d$, so that the  resonance state component comes only from $f_\alpha^S(\Omega)$ and becomes a symmetric Lorentzian.
Thus, the degree of the asymmetry (DA) for a particular resonance state $|\phi_\alpha\>$ is evaluated by the ratio of the third factors 
\begin{align}\label{DA}
{\rm DA}_\alpha\equiv {d \gamma_\alpha \over d\varepsilon_\alpha}\;,
\end{align}
 i.e. the tangent of the trajectories of Fig.\ref{Fig:ResonanceStates}.
The direction along the trajectory determines the direction of the asymmetry.
 

We gain deeper insight by comparing the resonance state component $f_\alpha(\Omega)$ with the ordinary Fano profile  
\begin{align}\label{FanoProfile}
f_{\rm F}(x)&={(x+q)^2\over x^2+1}=1+{q^2-1 \over x^2+1}+ {2q \, x \over x^2+1} \;.
\end{align}
In Eqs.(\ref{fSA}), with the definition 
\begin{align}\label{xdenine}
x\equiv {\Omega-\varepsilon_\alpha \over \gamma_\alpha}\;,
\end{align}
the symmetric and the antisymmetric parts of the resonance state component are rewritten as
\begin{subequations}\label{fSAx}
\begin{align}
f_\alpha^S(\Omega)&=\left({ \mu^2 T_{dc}^2 \over\pi \gamma_\alpha}\right) {1 \over x^2+1} {d\varepsilon_\alpha\over d E_d} \;, \\
f_\alpha^A(\Omega)&=\left({ \mu^2 T_{dc}^2 \over\pi \gamma_\alpha}\right) {x \over x^2+1}{d\gamma_\alpha \over d E_d} \;.
\end{align}
\end{subequations}
By comparing Eqs.(\ref{FanoProfile}) and (\ref{fSAx}), it follows that we  may evaluate the $q$-factor for a resonance state $|\phi_\alpha\>$ as
\begin{align}\label{Fanoq}
q={1\over {\rm DA}_\alpha}\left[ 1\pm\sqrt{1+{\rm DA}_\alpha^2} \right] \;,
\end{align}
where the sign is chosen according to the sign of $d\gamma_\alpha/dE_d$.
From this relation, we find
\begin{subequations}
\begin{align}
 \left|q\right| \to 1 &\text{ as }\left| DA_\alpha\right|\to \infty  \;, \\
\left| q\right|\to \infty  &  \text{ as }  DA_\alpha   \to 0\;.  
\end{align}
\end{subequations}
It should be noted that, since the relation Eq.(\ref{Fanoq}) is determined by solving the complex eigenvalue problem of the total Hamiltonian, our method enables us to rigorously determine the Fano $q$-factor based on  microscopic dynamics.

\begin{figure}[t]
\includegraphics[height=12cm,width=6cm]{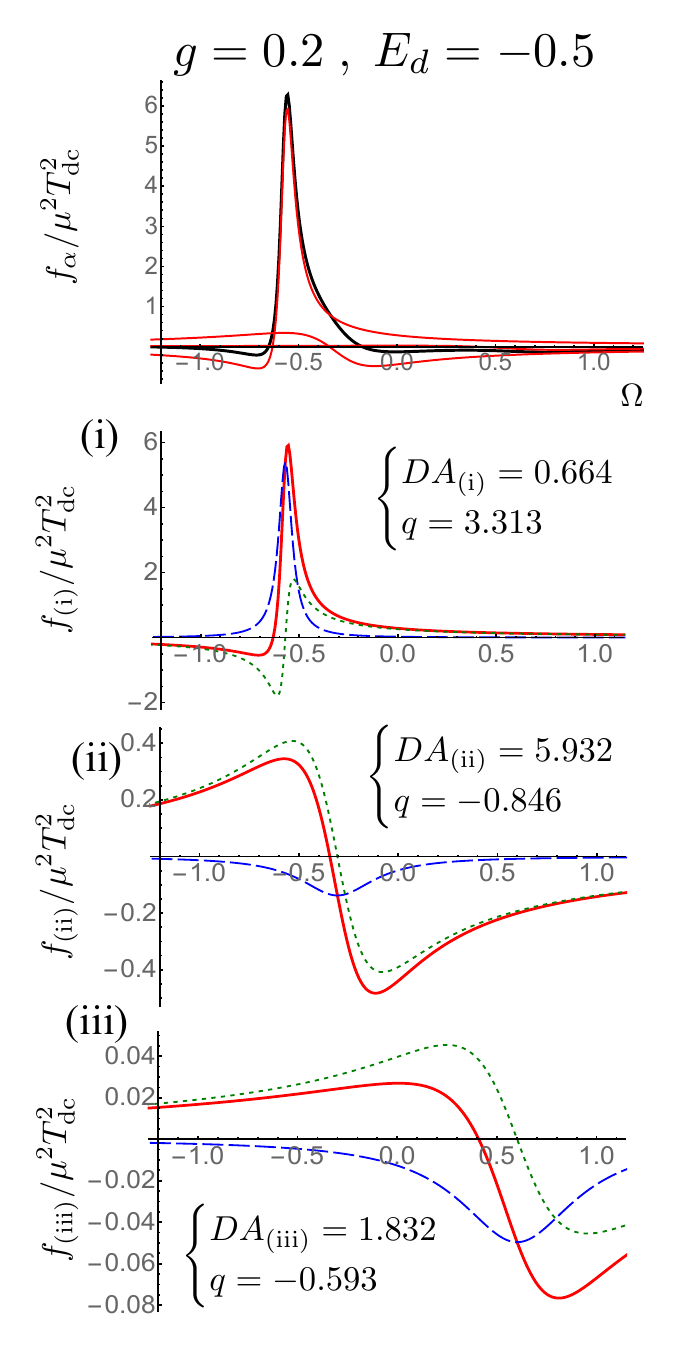}
\caption{(Color online) Resonant state components of the absorption spectrum for $g=0.2$ and $E_d=-0.5$ in Fig.\ref{Fig:absAll}.
In the top panel, the three resonance state components $f_\alpha(\Omega)$ are shown by the thin (red) lines, and the sum of them are depicted by the (black) thick line.
In the three lower panels, we have decomposed $f_\alpha(\Omega)$ into the symmetric $f^S_\alpha(\Omega)$ and antisymmetric $f^A_\alpha(\Omega)$ components shown by the (blue) dashed lines and (green) dotted lines, respectively.
The values of $DA_\alpha$ and the Fano $q$-factors are also shown in the inset.
}
\label{Fig:absFano}
\end{figure}

We show in Fig.\ref{Fig:absFano} the three resonance state components of the absorption spectrum for $g=0.2$ and $E_d=-0.5$ in Fig.\ref{Fig:absAll}.
The corresponding three resonance states are shown by the (red) filled circles in Fig.\ref{Fig:ResonanceStates}(c).
In the top panel, the three resonance state components $f_\alpha(\Omega)$ are shown by the thin (red) lines, and the sum of them are depicted by the (black) thick line.
In the lower panels of Fig.\ref{Fig:absFano}, we have decomposed $f_\alpha(\Omega)$ into the symmetric $f^S_\alpha(\Omega)$ and antisymmetric $f^A_\alpha(\Omega)$ parts shown by the (blue) dashed lines and (green) dotted lines, respectively, where $DA_\alpha$ and the Fano $q$-factors evaluated by Eq.(\ref{Fanoq}) are also shown in the inset.

We see that the main contribution comes from the resonance state component $f_{\rm (i)}(\Omega)$.
 This is because the maximum values of the spectral profile are given by $1/2\gamma_\alpha$ as mentioned above,  so that the resonance state with the smallest value of $\gamma_{\rm (i)}$ is mostly attributed to the spectrum.
The spectrum $f_{\rm (i)}(\Omega)$ exhibits a  sharp Fano profile with $DA_{\rm (i)}=0.664$ and $q=3.313$, while the antisymmetric parts overwhelms the symmetric parts for the nonanalytic resonance states (ii) and (iii) with large DA values. 

\begin{widetext}

\begin{figure}[t]
\includegraphics[height=12cm,width=14cm]{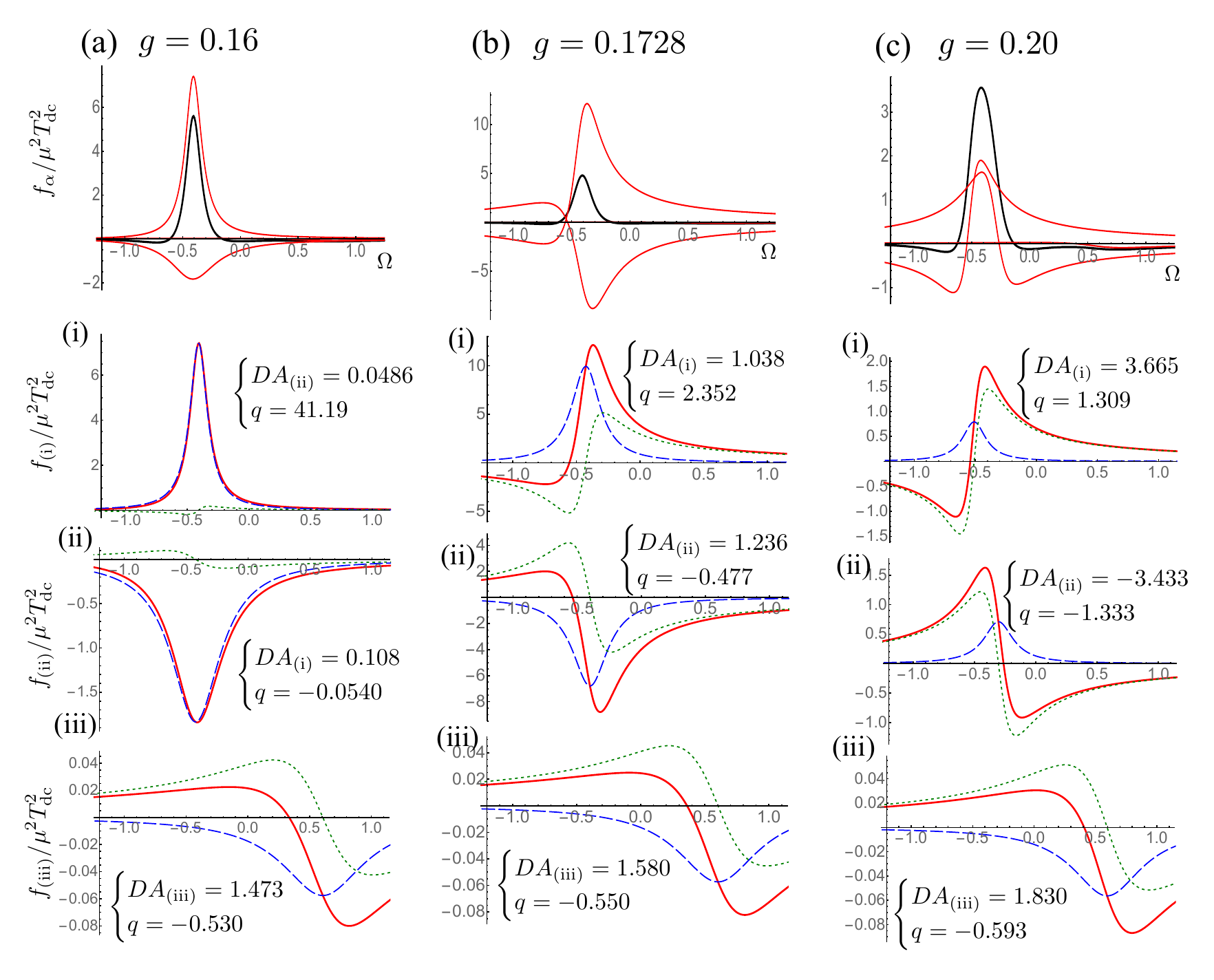}
\caption{(Color online) The resonance state components in the case of $E_d=-0.4$ for (a) $g=0.16$, (b) $g=0.1728$, and (c) $g=0.2$, corresponding to the case of $E_d=-0.4$ in Fig.\ref{Fig:absAll}.
The depiction of the spectrum decomposition is the same as in Fig.\ref{Fig:absFano}.
}
\label{Fig:absCompo}
\end{figure}
\end{widetext}


Next we study the difference of the absorption spectrum reflecting the trajectories of the complex eigenvalues shown in Fig.\ref{Fig:ResonanceStates}.
We show in Fig.\ref{Fig:absCompo} the resonance state components in the case $E_d=-0.4$ for (a) $g=-0.16$, (b) $g=-0.1728$, and (c) $g=-0.2$, corresponding to the case  $E_d=-0.4$ in Fig.\ref{Fig:absAll}.
The three resonance states for each cases are shown by the filled squares in Fig.\ref{Fig:ResonanceStates}.
The depiction of the spectrum decomposition  is the same as in Fig.\ref{Fig:absFano}.

In the weak coupling case (a) $g=0.16 < g_{\rm EP}$, where the repulsion of the trajectories occurs in parallel to the real axis, since $\gamma_{(i)}$ is much smaller than those for the other nonanalytic resonance states, the spectrum is mostly governed by the resonance state component $f_{\rm (i)}$.
Since ${\rm DA}_{\rm (i),(ii)}\simeq 0$  as seen from Fig.\ref{Fig:ResonanceStates}(a), the antisymmetric parts are very small, so that $f_{\rm (i),(ii)}(\Omega)$ show almost symmetric Lorentzian profiles.
The symmetric component $f_{\rm (ii)}$ is negative, because $d\varepsilon_{\rm (ii)}/dE_d <0$, as indicated by the arrow in Fig.\ref{Fig:ResonanceStates}(a).

On the other hand in the case of  strong coupling (c) $g=0.2 >g_{\rm EP}$, since the repulsion of the trajectories occurs in parallel to the imaginary axis as shown in Fig.\ref{Fig:ResonanceStates}, ${\rm DA}_{\rm (i),(ii)}$ becomes large, so that the antisymmetric part $f^A_{\rm (i),(ii)}$ overwhelms the symmetric part $f^S_{\rm (i),(ii)}$ as shown in Fig.\ref{Fig:absCompo}(c).
The direction of the asymmetric spectral profiles of $f^A_{\rm (i)}$ and $f^A_{\rm (ii)}$ are opposite.
 Because  $d \gamma_\alpha/d E_d$ has opposite sign between them as seen in Fig.\ref{Fig:ResonanceStates}(c), hence the spectral components cancel each other except around $\Omega\simeq -0.5$.
As a result, the resonance state component $\sum_\alpha f_\alpha(\Omega)$ shows a broad single Gaussian-type peak: Even though the absorption spectrum is similarly a single peak to Fig.\ref{Fig:absCompo}(a), its origin is very different.

For the case  (b) $g=0.1728 \simeq g_{\rm EP}$, it is striking that both the symmetric and antisymmetric components become very large and yet cancel each other between the resonance states (i) and (ii). (See the vertical scale of Fig.\ref{Fig:absCompo}(b).)
This reflects the fact that the normalization constant diverges at the EP, as mentioned above.
As a result of this (partial) cancellation,  the sum of the resonance components show a single peak.

\section{Discussion}\label{Sec:Discussion}
 
In this paper, we have  presented a new perspective on the absorption spectrum in terms of the complex spectral analysis.
We have studied the specific example of  the absorption spectrum of an impurity embedded in semi-infinite superlattice.
It is found that due to the boundary condition on the lattice, the self-energy has a strong energy dependence over the entire energy range of the continuum, which enhances the nonlinearity of  the eigenvalue problem of the effective Hamiltonian, yielding several nonanalytic resonance states  with respect to the coupling constant at $g=0$.

It has been revealed that the  overall spectral features are almost perfectly determined by the direct transitions to these discrete resonance states, reflecting the characteristic change in the complex energy spectrum of the total Hamiltonian. 
Even with only a single optical transition channel present, the absorption spectrum due to the transition to the resonance states, in general, takes an asymmetric Fano profile.
The asymmetry of the absorption spectrum is exaggerated for the transition to the nonanalytic resonance state. 
Since this is a genuine eigenstate of the total Hamiltonian, there is no ambiguity in the interpretation of the origin of the asymmetric profile of the absorption spectrum, avoiding the arbitrary interpretation based on  the quantum interference.

\begin{figure}
\begin{center}
\includegraphics[height=6cm,width=8cm]{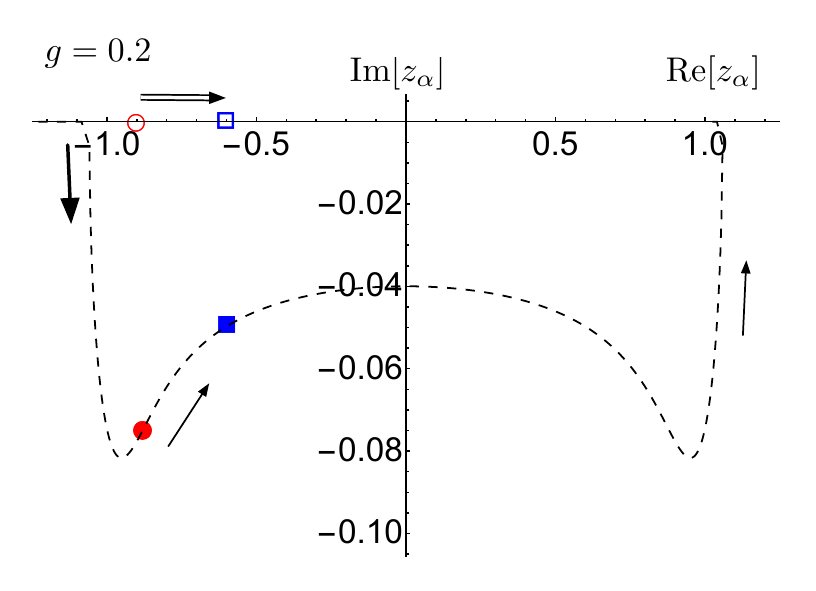}
\caption{(Color online) Resonance state solutions $z_\alpha$ of the infinite chain system for $g=0.2$ for a fixed value of  $E_d=-0.9$  (filled circle) and $E_d=-0.6$ (filled square) in the complex energy plane, where the horizontal and the vertical axes denote the real and the imaginary parts of the eigenvalues.
The open circle and square denote the position of the bare impurity energies $E_d$.
The dotted lines are the trajectory of the resonance state solutions with the change of $E_d$.
The arrows indicate the direction of the solutions along the trajectories as the bare energy increases as shown by the double arrows.
}
\label{Fig:TrajectoryInf}
\end{center}
\end{figure}

In order to illustrate the physical impact of the nonlinearity in the present system,  it is interesting to compare the present results with the absorption spectrum of the unbounded chain system, studied in Ref.\cite{Tanaka06PRB}, in terms of the resonant state representation.
For the unbounded  chain, the Hamiltonian is represented by
\begin{align}\label{HtkInf}
\hat{\cal H}  &= E_c |c\>\<c| + E_{d} | d \>\< d | + \int_{-\pi}^\pi  E_{k} | k \>\< k |  dk    \notag\\
&\quad +  \int_{-\pi}^\pi gV \left(   | k \rangle \langle d | + |d\>\<k|  \right)  dk\;,
\end{align}
where the energy dispersion $E_k$ is  the same as Eq.(\ref{Ek}). 
The important difference from Eq.(\ref{AppEq:H0V}) is that in this case the interaction potential does not depend on the wave number.
Then the effective Hamiltonian is given by $\hat{\cal H}_{\rm eff}(z)=E_d+g^2 \sigma^+(z)$, where the self-energy is given by
\begin{align}\label{SelfInf}
\sigma^+(z)= {1\over 2\pi} \int_{-\pi}^\pi dk {V^2 \over (z-E_k)^+} ={V^2\over \sqrt {z^2-1}} \;.
\end{align}
The dispersion equation $\eta^+(z)\equiv z-E_d-g^2\sigma^+(z)=0$ reduces to a fourth order polynomial equation, yielding a resonance state and an anti-resonance state in addition to the two bound states that are called as Persistent Bound States: PBS \cite{Garmon09,Tanaka06PRB}.\footnote{The definition for PBS is introduced in Ref.\cite{Garmon09}.}
We show in Fig.\ref{Fig:TrajectoryInf} the trajectory of the resonance state by the dotted line and the solutions for $E_d=-0.9$ and $E_d=-0.6$, by filled circle and filled square, respectively.
In contrast to the semi-infinite chain system, there is only a single resonance state in the infinite chain system, because the self-energy does not exhibit a strong energy dependence within the energy range of the continuum except for the band edges.

\begin{figure}[t]
\begin{center}
\includegraphics[height=11cm,width=7cm]{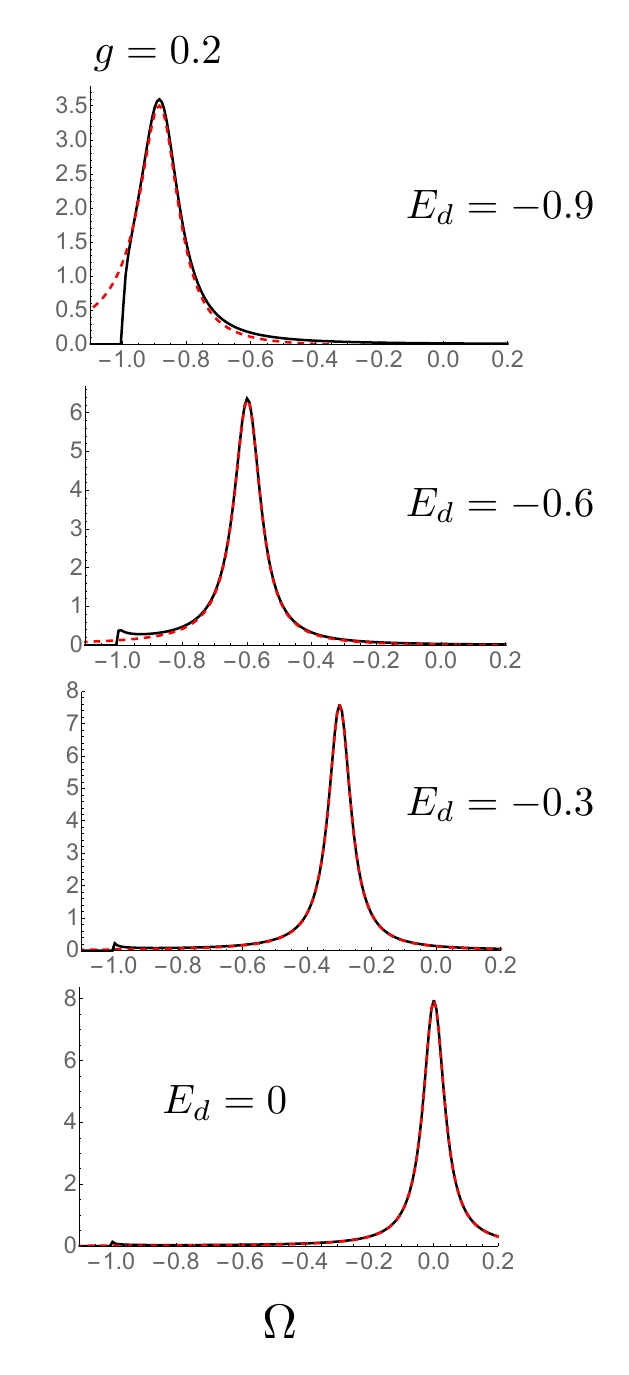}
\caption{(Color online) Absorption spectra $F(\Omega)$  for the infinite chain for $g=0.20$, where the spectra are divided by $\mu^2 T_{\rm dc}^2$: $F(\Omega)/\mu^2 T_{\rm dc}^2$.
In each panels, we show the absorption spectra for $E_d=-0.9, -0.6, -0.3,$ and $0$ from the top to bottom. 
The solid lines (black) represent $F(\Omega)$ and the dotted lines (red) represent the resonance component of Eq.(\ref{FoComSemi}).
}
\label{Fig:absInf}
\end{center}
\end{figure}

The absorption spectrum is calculated in terms of the complex eigenstate of the total Hamiltonian in a similar manner to the preceding section.
We have shown in Fig.\ref{Fig:absInf} the absorption spectra for $g=0.2$, where the bare impurity state energies $E_d$ are taken at $E_d=-0.9, -0.6, -0.3$, and 0.
It is seen that the overall spectral features shown by the (black) solid lines are perfectly reproduced by the resonance state components  (shown by the (red) dotted lines), just as in Fig.\ref{Fig:absAll}; Significant deviation appears only for the case  $E_d=-0.9$, where the self-energy changes due to the branch point effect, as explaind just below Eq.(\ref{Self}). 
In contrast to Fig.\ref{Fig:absAll}, however, the spectral profiles do not change  so much  with $E_d$: The single peak just shifts toward the higher energy side as $E_d$ increases, because the self-energy does not have strong energy dependence within the continuum, as mentioned above.

\begin{figure}[t]
\begin{center}
\includegraphics[height=7cm,width=6cm]{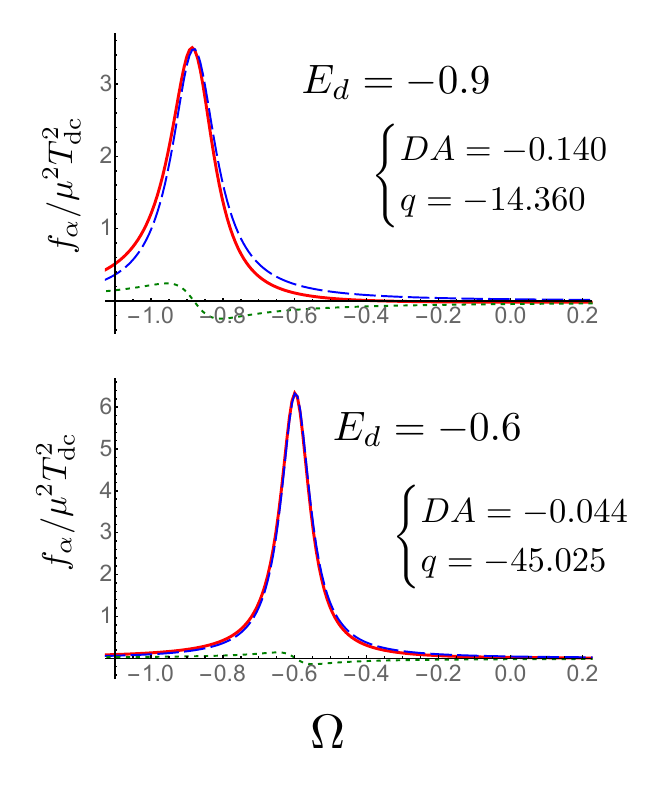}
\caption{(Color online) The resonance state components in the cases of $E_d=-0.9$ and $E_d=-0.9$  for $g=0.2$ corresponding to Fig.\ref{Fig:absInf}.
The depiction of the spectrum decomposition is the same as in Fig.\ref{Fig:absFano}.
}
\label{Fig:absCompoInf}
\end{center}
\end{figure}

In order to confirm that the absorption spectrum of the transition to the resonance states takes an asymmetric Fano shape, even with a single optical transition channel, we decompose the resonance state component for $E_d=-0.9$ and $E_d=-0.6$ into  symmetric and antisymmetric parts in Fig.\ref{Fig:absCompoInf}.
We found that the degree of the asymmetry is always non-zero although it is quite small, therefore the absorption spectrum exhibits a Fano-type asymmetry.
However, compared to the semi-infinite lattice case, the degree of the asymmetry is very small so that the absorption spectrum takes an almost symmetric Lorentzian shape.

The present method for interpreting the absorption spectrum in terms of the direct transition to the discrete resonance states is an extension of  Bohr's idea for   quantum jumps between  discrete states of  matter under optical transitions.
In the usual picture, the spectrum due to the quantum jump just exhibits a symmetric Lorentzian profile, whose peak position and width are determined by the excitation energy and the decay rate, i.e. the real and imaginary parts of the complex eigenvalues, respectively.
What we have shown here is that the optical spectrum due to the quantum jump between the resonance states can cause much richer spectral features, representing not only their complex eigenvalues but also the peculiar features of the wave functions belonging to the extended Hilbert space.
Therefore,  we hope that the present method can be applied to give a new understanding for stationary spectroscopies, such as resonance fluorescence, four-wave mixing, etc., in the frequency domain, but also for time-resolved spectroscopies \cite{Gruson16Science,Kaldun16Science,Misochko15JETP}.


\acknowledgements

We are very grateful Dr. Tomio Petrosky for many valuable discussions.
We also thank K. Mizoguchi and Y. Kayanuma for fruitful comments.
This work was partially supported by JSPS Grant-in-Aid for Scientific Research No.16H04003, 16K05481, and 17K05585.
　　　　　　　　　　　　　　　　　　　　　　　　　　　　　　　　　　　　
\appendix　
\section{Absorption spectrum in terms of Green's function method}\label{App:Green}

In this section, we briefly review the Green's function method to evaluate the absorption spectrum Eq.(\ref{Fo}).
Defining the resolvent operator as
\begin{align}\label{AppEq:resolvent}
\hat G(z)\equiv {1\over z-\hat H}={1\over z-\hat H_0-g\hat V} \;,
\end{align}
where $\hat H_0$ and $g\hat V$ are given by Eq(\ref{AppEq:H0V}).
With use of the Dyson's equation, we have the relations
\begin{align}\label{Green}
G_{dd}(z)&\equiv \<d|\hat G(z)|d\>={1\over z-E_d}+{1\over z-E_d} \int dk g V_k G_{kd} \;, \\
G_{kd}(z)&\equiv \<k|\hat G(z)|d\>={1\over z-E_k}gV_k G_{dd} \;,
\end{align}
where $G_{ij}(z)$ is an element of the resolvent.
It immediately follows from Eqs.(\ref{Green}) that we obtain
\begin{align}\label{AppEq:Gdd}
G_{dd}(z)={1\over z-E_d-g^2 \Sigma^+(z)} \;.
\end{align}
Therefore, the absorption spectrum is obtained by substituting Eq.(\ref{AppEq:Gdd}) into Eq.(\ref{Fo}) as
\begin{align}
F(\Omega)=-{\mu^2T_{dc}^2\over\pi}{g^2 {\rm Im}\Sigma^+(\Omega) \over \left(\Omega-E_d-g^2{\rm Re}\Sigma^+(\Omega) \right)^2+g^4 \left({\rm Im}\Sigma^+(\Omega)\right)^2} \;,
\end{align}
where the self-energy is defined by Eq.(\ref{Self}).

\section{Complex eigenvalue problem with the projection method}\label{App:EVP}

In this section we briefly summarize the complex eigenvalue problem with use of  the BWF projection method.
One could refer to the literatures for details\cite{Petrosky91Physica,Rotter09JPhysA,Tanaka16PRA}.

First, we consider the right-eigenstate for the discrete resonance state.
\begin{align}\label{AppEq:rightEVP}
\hat H|\phi_\alpha\>=z_\alpha|\phi_\alpha\> \;.
\end{align}
The  application of the projection operators given in Eq.(\ref{Projections}) to the above leads to
\begin{subequations}\label{AppEq:Projection_d}
\begin{align}
\hat P^{(d)} \hat H_0\hat P^{(d)} |\phi_\alpha\>+\hat P^{(d)} g\hat V\hat Q^{(d)} |\phi_\alpha\>&=z_d\hat P^{(d)}|\phi_\alpha\>  \;,\\
\hat Q^{(d)} g\hat V\hat P^{(d)} |\phi_\alpha\>+\hat Q^{(d)} \hat H\hat Q^{(d)} |\phi_\alpha\>&=z_d\hat Q^{(d)}|\phi_\alpha\> \;.
\end{align}
\end{subequations}
The $\hat Q^{(d)}|\phi_\alpha\>$ is solved in Eq.(\ref{AppEq:Projection_d}b) as
\begin{align}\label{AppEq:Qphi}
\hat Q^{(d)}|\phi_\alpha\>={1\over z_\alpha-\hat Q^{(d)} \hat H\hat Q^{(d)}}\hat Q^{(d)}\hat H\hat P^{(d)}|\phi_\alpha\> \;,
\end{align}
which is substituted to Eq.(\ref{AppEq:Projection_d}a),  we then have the eigenvalue problem of the effective Hamiltonian Eq.(\ref{HeffEVP}), where the effective Hamiltonian is expressed by
\begin{subequations}
\begin{align}
\hat H_{\rm eff}(z)&=\hat P^{(d)}\hat H_0  P^{(d)} \notag\\
&\quad+  P^{(d)}\hat V\hat Q^{(d)} {g^2\over z-\hat Q^{(d)} \hat H \hat Q^{(d)}} \hat Q^{(d)} \hat V  P^{(d)}  \\
&=E_d+g^2\Sigma^+(z)\;.
\end{align}
\end{subequations}
The discrete resonance state eigenvalues are obtained as the solutions of  the dispersion equation Eq.(\ref{DispersionSemi}).
The corresponding resonance state is obtained by adding the $\hat Q^{(d)}$ component as
\begin{align}\label{AppEq:rightES}
|\phi_\alpha\>&=\hat P^{(d)}|\phi_\alpha\>+\hat Q^{(d)}|\phi_\alpha\>  \\
&=\<d|\phi_\alpha\>\left( |d\>+g   \int_0^\pi dk {V_k \over (z-E_k)^+_{z=z_\alpha}} |k\>\right) 
\end{align}
where we have used Eq.(\ref{Projections}).

The left-eigenstate problem
\begin{align}
\<\tilde\phi_\alpha|\hat H=z_\alpha\<\tilde\phi_\alpha|
\end{align}
is similarly solved by applying the projection operators from the right.

Next we solve for the continuous eigenstate.
For this purpose, we choose the projection operators as
\begin{align}
\hat P^{(k)}=|k\>\<k| \;, \hat Q^{(k)}=1-\hat P^{(k)} \;.
\end{align}

The effective Hamiltonian for $k$-space is given by
\begin{align}\label{AppEq:Heffk}
\hat H_{\rm eff}^{(k)}(z)&=\hat P^{(k)}\hat H_0  P^{(k)} \notag\\
&\quad+  P^{(k)}\hat V\hat Q^{(k)} {g^2\over z-{\cal H}^{(k)}} \hat Q^{(k)} \hat V  P^{(k)}   \;,
\end{align}
where we have denoted
\begin{equation}
{\cal H}^{(k)}\equiv \hat Q^{(k)} \hat H \hat Q^{(k)} \;.
\end{equation}
This is represented by
\begin{subequations}
\begin{align}
{\cal H}^{(k)}&=E_d|d\>\<d|+\sum_{k'(\neq k)} E_{k'}|k'\>\<k'|  \notag\\
&+{2 \over\sqrt {N}} g \sum_{k'(\neq k)}\sin k' \left( |d\>\<k'|+|k'\>\<d|\right) \\
&\equiv {\cal H}_0^{(k)} +  {\cal V}^{(k)}  \;.
\end{align}
\end{subequations}

Then the matrix elements in the second term of Eq.(\ref{AppEq:Heffk}) are represented by
\begin{align}\label{AppEq:HeffMat}
 \<k| \hat V \hat Q^{(k)}{1\over  z_k-{\cal H}^{(k)}  } Q^{(k)} \hat V |k\> = {4\over N}(\sin k)^2  g^2 G_{dd}^{(k)}(z_k)
\end{align}
where the Green's function in terms of the $k$-state is given by
\begin{align}
G_{dd}^{(k)}(z_k)=  \<d|{1\over z_k-{\cal H}^{(k)}}  |d\>\;.
\end{align}

By using the Dyson equation, we have the relations
\begin{subequations}\label{AppEq:Greenk}
\begin{align}
G_{dd}^{(k)}(z)&={1\over z-E_d}+{1\over z-E_d}{2g \over\sqrt N}\sum_{k'(\neq k)}\sin k' G_{k' d}^{(k)}(z) \;, \\
G_{k'd}^{(k)}(z)&={1\over z-E_{k'}}{2 \sin k' \over\sqrt N}  g G_{dd}^{(k)}(z) \;.
\end{align}
\end{subequations}
Substitution of Eq.(\ref{AppEq:Greenk}b) into (\ref{AppEq:Greenk}a) yileds
\begin{align}\label{AppEq:Gddk}
G_{dd}^{(k)}(z)= {1\over\eta^+(z)}  \;,
\end{align}
where $\eta^+(z)$ is given in Eq.(\ref{DispersionSemi}).

Using Eqs.(\ref{AppEq:Heffk}), (\ref{AppEq:HeffMat}), and   (\ref{AppEq:Gddk}), the eigenvalue problem of the effective Hamiltonian $\hat H_{\rm eff}^{(k)}(z)$ reads
\begin{align}
\hat H_{\rm eff}^{(k)}(z_k)\hat P^{(k)}|\phi_k\>&=\left[ E_k+  {4(\sin k)^2 \over N} {g^2 \over\eta^+(z_k)} \right]\hat P^{(k)}|\phi_k\> \notag\\
&=z_k \hat P^{(k)}|\phi_k\> \;.
\end{align}
We find that $z_k=E_k$ in the limit $N\to\infty$.

The right-continuous eigenstate for the wave number  $k$ is given by adding the $\hat Q^{(k)}$ component
\begin{align}\label{AppEq:phik}
|\phi_k\>&=\hat P^{(k)}|\phi_k\>+\hat Q^{(k)}|\phi_k\> \notag \\
&=\left[|k\>+\hat Q^{(k)} {1\over E_k -{\cal H}^{(k)} } \hat Q^{(k)} \hat V |k\> \right]\<k|\phi_k\> \;.
\end{align}
The second term is written by
\begin{subequations}\label{AppEq:QkCompo}
\begin{align}
&\hat Q^{(k)} {1\over E_k -{\cal H}^{(k)} } \hat Q^{(k)} \hat V |k\>  \nonumber\\
&= |d\>\<d|{1\over E_k - {\cal H}^{(k)}  }|d\>\<d| \hat V |k\> \nonumber\\
&+\sum_{k'(\neq k)}|k'\>\<k'|{1\over E_k - {\cal H}^{(k)}  }|d\>\<d| \hat V |k\> \notag \\
&={2gV \sin k  \over \sqrt N}\left[ |d\>  G_{dd}^{(k)}(E_k)+\sum_{k'(\neq k)} |k'\>  G_{k'a}^{(k)}(E_k)  \right]  \;.
\end{align}
\end{subequations}
Substituting Eq.(\ref{AppEq:QkCompo}) into Eq.(\ref{AppEq:phik}), we obtain Eq.(\ref{phikR}).
The left-eigenstate for the continuous state is obtained in the same way.

\section{Dispersion equation in a form of a polynomial}\label{AppSec:Dispersion}

In this section we reduce the dispersion equation Eq.(\ref{DispersionSemi}) to a $2n_d$-th order polynomial equation.
Using a binomial expansion, it is written as  
\begin{widetext}
\begin{align}\label{AppEq:DispExpansion1}
&z - E_d +  \frac{g^2}{ \sqrt{z^2 - 1} } \sum_{m=0}^{n_d {\!} -1 } \binom{2 n_d {\!} }{ 2m+1 } ( -z )^{2n_d {\!} - (2m + 1) } \Bigl( \sqrt{ z^2 - 1 }  \Bigr)^{2m+1} \notag\\
&\qquad  =\frac{g^2}{ \sqrt{z^2 - 1} } \biggl[ 1 - \sum_{m=0}^{n_d {\!} } \binom{2 n_d {\!} }{ 2m } ( -z )^{2n_d {\!} - 2m } \Bigl( \sqrt{ z^2 - 1 }  \Bigr)^{2m}  \biggr] \;.
\end{align}
Using the identities
\begin{subequations}\label{AppEq:Expansion}
\begin{align}
&\sum_{m=0}^{n_d {\!} -1 } \binom{2 n_d {\!} }{ 2m+1 } ( -z )^{2n_d {\!} - (2m + 1) } \Bigl( \sqrt{ z^2 - 1 }  \Bigr)^{2m+1} =  \frac{1}{2} \biggl\{ \Bigl(  -z + \sqrt{z^2 - 1}  \Bigr)^{2n_d {\!} } - \Bigl(  -z - \sqrt{z^2 - 1}  \Bigr)^{2n_d {\!} } \biggr\} \;,\\
&\sum_{m=0}^{n_d {\!} } \binom{2 n_d {\!} }{ 2m } ( -z )^{2n_d {\!} - 2m } \Bigl( \sqrt{ z^2 - 1 }  \Bigr)^{2m}  = \frac{1}{2} \biggl\{ \Bigl(  -z + \sqrt{z^2 - 1}  \Bigr)^{2n_d {\!} } + \Bigl(  -z - \sqrt{z^2 - 1}  \Bigr)^{2n_d {\!} } \biggr\} \;,
\end{align}
\end{subequations}
Eq.(\ref{AppEq:DispExpansion1}) reads
\begin{align}\label{AppEq:zEd}
&z - E_d +  \frac{g^2}{ 2\sqrt{z^2 - 1} } \biggl[ \Bigl(  -z + \sqrt{z^2 - 1}  \Bigr)^{2n_d {\!} } - \Bigl(  -z - \sqrt{z^2 - 1}  \Bigr)^{2n_d {\!} } \biggr]  \notag\\
&\qquad =\frac{g^2}{ 2\sqrt{z^2 - 1} } \biggl[ 2 - \Bigl(  -z + \sqrt{z^2 - 1}  \Bigr)^{2n_d {\!} } + \Bigl(  -z - \sqrt{z^2 - 1}  \Bigr)^{2n_d {\!} }  \biggr] \;.
\end{align}

Taking the square of Eq.(\ref{AppEq:zEd}) and using Eq.(\ref{AppEq:Expansion}b) again, we obtain the $2n_d$-th order polynomial equation:
\begin{align}
&(z - E_d)^2 +  2 g^2 (z - E_d)  \biggl\{ \sum_{m=0}^{n_d {\!} -1 } \binom{2 n_d {\!} }{ 2m+1 } ( -z )^{2n_d {\!} - (2m + 1) } \Bigl(  z^2 - 1  \Bigr)^{m} \biggr\} \notag\\
&\qquad + 2 g^4 \biggl\{ \sum_{m=0}^{n_d {\!} -1} z^{2m} + \sum_{m=1}^{n_d {\!} } \binom{2 n_d {\!} }{ 2m } ( -z )^{2n_d {\!} - 2m } \Bigl(  z^2 - 1   \Bigr)^{m-1}  \biggr\} = 0.
\end{align}

\end{widetext}
　　　　　　　　　　　　　　　　　　　　　　　　　　　　　　　
　　　　　　　　　　　　　　　　　　　　　　　　　　　　　　　　　　　　　　　　　
　　　　　　　　　　　　　　　　　　　　　　　　　　　　　　　　　　　　　　

%


\end{document}